\documentclass[twocolumn,showpacs,showkeys,preprintnumbers]{revtex4}
\usepackage{amssymb}

\usepackage{amsmath}
\usepackage{hyperref}
\usepackage{graphicx}

\input{tcilatex}

\begin{document}

\title{The Entanglement in Anisotropic Heisenberg $XYZ$ Chain with
inhomogeneous magnetic field}
\author{G.H. Yang, W.B. Gao, L. Zhou, H.S.Song }
\affiliation{Department of Physics, Dalian University of Technology, Dalian 116023, PR
China}

\begin{abstract}
The thermal entanglement of a two-qubit anisotropic Heisenberg $XYZ$ chain
under an inhomogeneous magnetic field $b$ is studied. It is shown that when
inhomogeneity is increased to certain value, the entanglement can exhibit a
larger revival than that of less values of $b$. The property is both true
for zero temperature and a finite temperature. The results also show that
the entanglement and critical temperature can be increased by increasing
inhomogeneous exteral magnetic field.
\end{abstract}

\maketitle

\address{Department of Physics, Dalian University of Technology,}

\section{Introduction}

As a valuable resource in quantum information and quantum computation [1],
quantum entanglement has attracted numerous attention over past decade
years. In order to realize the quantum information process, a great deal of
effort has been devoted to study and characterize the entanglement in solid
state systems [2,3]. A typical example in solid system is the spin chains
which are the natural candidates for the realization of the entanglement
compared with the other physics systems. The Heiserberg model can describe
interaction of qubits not only in solid physical systems but also in many
other systems such as quantum dots [4] and nuclear spin [5]; therefore,
there are numerous studies on the Heisenberg models[6-13]. It is turned out
that the critical magnetic field $B_{c}$ is decreased with the increase of
the anisotropic parameter $\gamma $ but the critical temperature $T_{c}$ is
improved. When B crosses $B_{c}$ the concurrence C drops suddenly and then
undergoes a ''revival'' for sufficiently large $\gamma $ in Ref [14], but it
only discussed the uniform magnetic field case. As we all know, in any solid
state construction of qubits, there is always the possibility of
inhomogeneous Zeeman coupling [15,16]. Quite recently the effect of
inhomogeneous magnetic field on the thermal entanglement of an isotropic
two-qubit $XXX$ spin system has been studied in Ref [17]. Including
interaction of Z-component $J_{z}$, Ref [18] also studied the effect of
inhomogeneous magnetic field on the thermal entanglement in a two-qubit
Heisenberg $XXZ$ spin chain.

However, the entanglement for a $XYZ$ spin model under an inhomogeneous
magnetic field has not been discussed. Therefore, in this paper we
investigate the influence of an external inhomogeneous magnetic field on the
entanglement of a two-qubit Heisenberg $XYZ$ system at thermal equilibrium.
Our studies show that the inhomogenerous extral magnetic field can make the
revival entanglement larger,improve the critical temperature, and enhance
entanglement.

\section{Theoretical treatment and results}

The Heisenberg Hamiltonian of a N-qubit anisotropic Heisenberg $XYZ$ model
under an inhomogeneous magnetic field is

\begin{eqnarray}
H &=&\frac{1}{2}\sum_{i=1}^{N}[J_{x}\sigma _{i}^{x}\sigma
_{i+1}^{x}+J_{y}\sigma _{i}^{y}\sigma _{i+1}^{y}+J_{z}\sigma _{i}^{z}\sigma
_{i+1}^{z}  \notag \\
&&+(B+b)\sigma _{i}^{z}+(B-b)\sigma _{i+1}^{z}],  \label{eq1}
\end{eqnarray}%
where ($\sigma _{i}^{x}$, $\sigma _{i}^{y}$, $\sigma _{i}^{z}$) are the
vector of Pauli matrices and $J_{j}$ ($j=x,y,z$) is the anisotropic coupling
coefficient between the nearest two spins. The parameter $J_{j}>0$ means
that the chain is antiferromagnetic, and ferromagnetic for $J_{j}<0$. The
magnetic fields on the nearest-neighbor two qubits are $B-b$ and $B+b$,
respectively, the value of $b$ controls the degree of inhomogeneity.

For a spin system in equilibrium at temperature $T$, the density matrix is $%
\rho =(1/Z)\exp (-H/k_{B}T)$, where $H$ is the Hamiltonian of this system, $%
Z $ is the partition function and $k_{B}$ is the Boltzmann constant. Usually
we write $k_{B}=1$. For a two-qubit system the thermal entanglement can be
measured by the concurrence $C$ which can be calculated with the help of $%
Wootters^{,}$ formula\cite{wooters} $C=\max (0,2\max {\lambda _{i}}%
-\sum_{i=1}^{4}\lambda _{i})$, where $\lambda _{i}$ is the square roots of
the eigenvalues of the matrix 
\begin{equation}
R=\rho (\sigma _{1}^{y}\otimes \sigma _{2}^{y})\rho ^{\ast }(\sigma
_{1}^{y}\otimes \sigma _{2}^{y}),  \label{eq2}
\end{equation}%
where the asterisk indicates complex conjugation, the concurrence $C$ ranges
from zero to one.

Consider now the Hamiltonian $H$ for the anisotropic two-qubit Heisenberg $%
XYZ$ chain in an inhomogeneous magnetic field. The Hamiltonian can be shown
as 
\begin{eqnarray}
H &=&J(\sigma _{1}^{+}\sigma _{2}^{-}+\sigma _{1}^{-}\sigma
_{2}^{+})+J\gamma (\sigma _{1}^{+}\sigma _{2}^{+}++\sigma _{1}^{-}\sigma
_{2}^{-})+\frac{J_{z}}{2}\sigma _{1}^{z}\sigma _{2}^{z}  \notag \\
&&+\frac{(B+b)}{2}\sigma _{1}^{z}+\frac{(B-b)}{2}\sigma _{2}^{z},  \label{3}
\end{eqnarray}%
where $J=\frac{(J_{x}+J_{y})}{2}$, $\gamma =\frac{J_{x}-J_{y}}{J_{x}+J_{y}}$
and $\sigma ^{\pm }=\frac{1}{2}(\sigma ^{x}\pm {i}\sigma ^{y})$. Among these
parameters $\sigma ^{\pm }$ are raising and lowering operators respectively
and $\gamma $ $(0<\gamma <1)$ measures the anisotropy in the $XY$ plane. In
the standard basis $\{|00\rangle ,|01\rangle ,|10\rangle ,|11\rangle \}$,
the Hamiltonian can be expressed as 
\begin{equation*}
H=\left( 
\begin{array}{cccc}
\frac{J_{z}}{2}+B & 0 & 0 & J\gamma \\ 
0 & -\frac{J_{z}}{2}+b & J & 0 \\ 
0 & J & -\frac{J_{z}}{2}-b & 0 \\ 
J\gamma & 0 & 0 & \frac{J_{z}}{2}-B%
\end{array}%
\right) .
\end{equation*}

The eigenvectors and eigenvalues of $H$ are easily obtained as following
forms $H|\psi^{\pm}\rangle=(-\frac{J_z}{2}\pm\xi)|\psi^{\pm}\rangle$ and $%
H|\Sigma^{\pm}\rangle=(\frac{J_z}{2}\pm\eta)|\Sigma^{\pm}\rangle$, with the
eigenstates are $|\psi^{\pm}\rangle=N^{\pm}[\frac{(b\pm\xi)}{J}%
|01\rangle+|10\rangle]$ and $|\Sigma^{\pm}\rangle=M^{\pm}[\frac{(B\pm\eta)}{%
J\gamma}|00\rangle+|11\rangle]$, where $\eta=\sqrt{B^2+J^2\gamma^2}$ and $%
\xi=\sqrt{b^2+J^2}$, the normalization constants are $N^{\pm}=1/\sqrt{%
1+(b\pm\xi)^2/J^2}$ and $M^{\pm}=1/\sqrt{1+(B\pm\eta)^2/J^2\gamma^2}$. One
can notice that these four eigenstates are all entanglement states when $%
J\neq0$ so it means that entanglement exists for both antiferromagnetic $%
(J>0)$ and ferromagnetic $(J<0)$ cases.

Now we can calculate the thermal entanglement of this system. The density
matrix can be written as the following form 
\begin{equation*}
\rho =\left( 
\begin{array}{cccc}
\mu _{+} & 0 & 0 & \upsilon \\ 
0 & \omega _{1} & z & 0 \\ 
0 & z & \omega _{2} & 0 \\ 
\upsilon & 0 & 0 & \mu _{-}%
\end{array}%
\right) .
\end{equation*}%
The square roots of the eigenvalues of the matrix $R$ are 
\begin{equation}
\lambda _{1,2}=|{\sqrt{\mu _{+}\mu _{-}}\pm \upsilon }|,\lambda _{3,4}=|{%
\sqrt{\omega _{1}\omega _{2}}\pm {z}}|.
\end{equation}%
The exact values of these nonzero matrix elements can be obtained by knowing
the spectrum of $H$. We obtain 
\begin{equation}
\mu _{+}=\frac{1}{Z}e^{-\frac{J_{z}}{2}\beta }[\cosh (\eta \beta )-\frac{%
\beta }{\eta }\sinh (\eta \beta )]  \notag \\
,
\end{equation}%
\begin{equation}
\mu _{-}=\frac{1}{Z}e^{-\frac{J_{z}}{2}\beta }[\cosh (\eta \beta )+\frac{%
\beta }{\eta }\sinh (\eta \beta )]  \notag \\
,
\end{equation}%
\begin{equation}
z=-\frac{J}{Z\xi }e^{\frac{J_{z}}{2}\beta }\sinh (\xi \beta )  \notag \\
,
\end{equation}%
\begin{equation}
\omega _{1}=\frac{1}{Z}e^{\frac{J_{z}}{2}\beta }[\cosh (\xi \beta )-\frac{b}{%
\xi }\sinh (\xi \beta )]  \notag \\
,
\end{equation}%
\begin{equation}
\omega _{2}=\frac{1}{Z}e^{\frac{J_{z}}{2}\beta }[\cosh (\xi \beta )+\frac{b}{%
\xi }\sinh (\xi \beta )]  \notag \\
,
\end{equation}%
\begin{equation}
\upsilon =-\frac{J\gamma }{Z\eta }e^{-\frac{J_{z}}{2}\beta }\sinh (\eta
\beta ),
\end{equation}%
where $Z$ is the partition function given by 
\begin{equation}
Z=tre^{-{\beta }H}=2[e^{-\frac{J_{z}}{2}\beta }\cosh (\eta \beta )+e^{\frac{%
J_{z}}{2}\beta }\cosh (\xi \beta )].
\end{equation}%
%
%
%
%
%
%
Thus from Eq.(4)-(6) we find that 
\begin{equation}
\lambda _{1,2}=\frac{1}{Z}e^{-\frac{J_{z}}{2}\beta }|\sqrt{1+\frac{%
J^{2}\gamma ^{2}}{\eta ^{2}}\sinh (\eta \beta )^{2}}\mp \frac{J\gamma }{\eta 
}\sinh (\eta \beta )|  \notag \\
,
\end{equation}%
\begin{equation}
\lambda _{3,4}=\frac{1}{Z}e^{\frac{J_{z}}{2}\beta }|\sqrt{1+\frac{J^{2}}{\xi
^{2}}\sinh (\xi \beta )^{2}}\mp \frac{J}{\xi }\sinh (\xi \beta )|.
\end{equation}%
We can calculate the thermal entanglement which is measured by the
definition of the concurrence. One can notice that Eq.(7) are the same as
the case of $b=0$ $[20]$. The concurrence which is directly determined by $%
\lambda _{i}(i=1,2,3,4)$ is invariant under the substitutions $\gamma
\longrightarrow -\gamma $ and $J\longrightarrow -J$ but the eigenvalue is
variant with the substitution $J_{z}\longrightarrow -J_{z}$ so that the
concurrence is also different. We will take the cases $J>0$ and $0\leq
\gamma \leq 1$ into account in this paper.

From Eq.(7) and the definition of the concurrence, we derive the $C$ for $%
T=0 $ as following 
\begin{equation}
C(T=0)=%
\begin{cases}
\frac{J\gamma }{\eta } & \xi <\eta -J_{z}, \\ 
|\frac{J\gamma }{\eta }-\frac{J}{\xi }|/2 & \xi =\eta -J_{z}, \\ 
\frac{J}{\xi } & \xi >\eta -J_{z}.%
\end{cases}%
\end{equation}%
Although the parameters $J$, $\gamma $, $\eta $ and $\xi $ are independent
of $J_{z}$ in the case of two interacting qubits, the value of $J_{z}$ \ is
very important in determining the point of the piecewise function so that it
can play role in the pairwise entanglement. We will show it in Fig.1.

\begin{figure}[tbp]
\includegraphics[width=2.6in, height=2.4in]{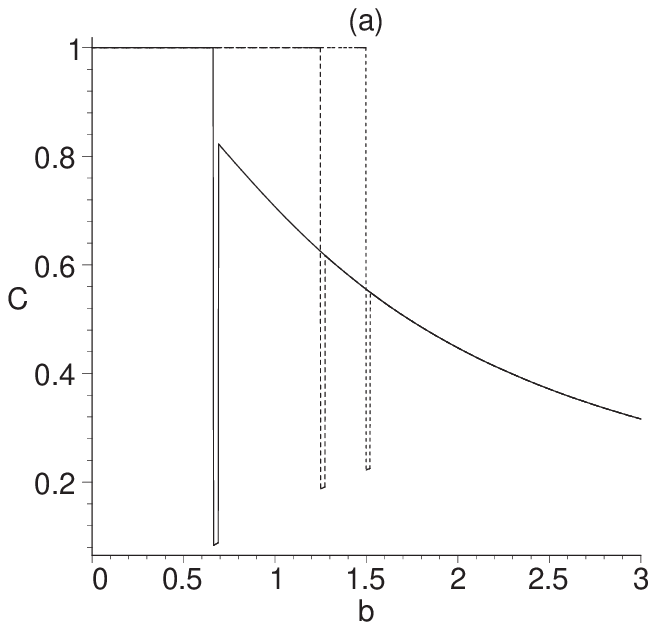} %
\includegraphics[width=2.60in, height=2.4in]{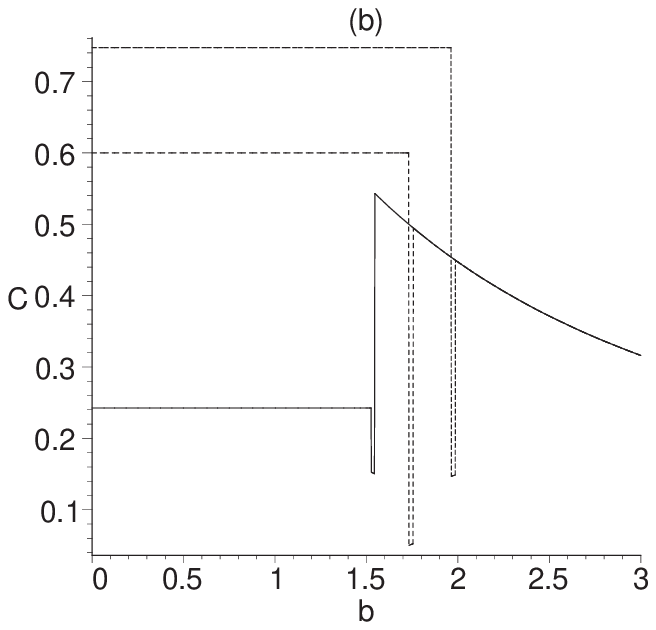} %
\includegraphics[width=2.6in, height=2.4in,]{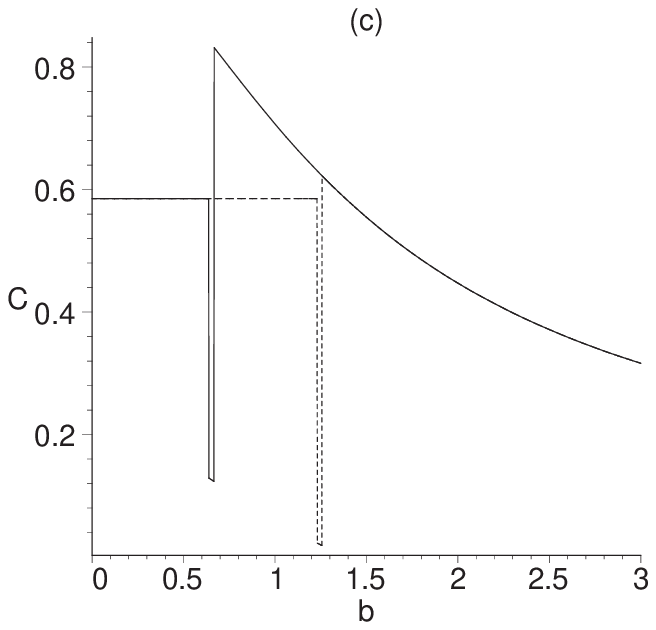}
\caption{The concurrence in the $XYZ$ model is plotted vs b for various
values of anisotropy parameter. (a):From left to right $\protect\gamma =0.2$
, $0.6$ ,$0.9,$ respectively, where $B=0$ and $J_{z}=-1$. (b): From bottom
to up $\protect\gamma =0.2,0.6,0.9,$ respectively, $B=0.8$ and $J_{z}=-1$.
(c) gives the concurrence with different values of $J_{z}$: $J_{z}=-0.2$
(solid line), $J_{z}=-0.6$ (dashed line) with $\protect\gamma =0.6$ and $%
B=0.8$. For all plotted $T=0$ and $J=1$.}
\end{figure}

The concurrence $C$ as a function of $b$ at $T=0$ for three values of $%
\gamma $ are given in Fig.1. Consider first the nonuniform magnetic field ($%
B=0$) case [Fig. 1(a)]. With increasing $b$, the concurrence $C$ is
initially constant and equal to its maximal value 1. It then drops suddenly
as a critical value $b_{c}$ is reached, where the critical inhomogeneous
magnetic field $b_{c}$ is given by $\xi =\eta -J_{z}$ or $b_{c}=\sqrt{(\eta
-J_{z})^{2}-J^{2}}$. At the critical point ($T=0$, $b=b_{c}$), the
entanglement becomes a nonanalytic function of $b$ and a quantum phase
transition occurs, but for $b>b_{c}$ the concurrence $C$ undergoes a revival
before decreasing to zero. The critical inhomogeneous magnetic field $b_{c}$
is increased by increasing the anisotropy parameter $\gamma $, which means
the region that entanglement keeps its maximum value is broaden with the
increasing $\gamma $. However, for $B\neq 0$ case, entanglement exhibits
novel property shown in Fig. 1(b). For certain relationship of the
parameters, the value of revival can be larger than its value before
dropping, for example, Fig1.(b) $\gamma =0.2$ curve. In addition, the region
of entanglement keeping constant and entanglement values is increased with
the increasing $\gamma $. In Fig.1 (c), we further show the role of $J_{z}$
in existing larger revival. One can see clearly the values of $J_{z}$
broaden the region of entanglement keeping constant values. The most
important is that only when $J_{z}$ achieve certain values, the value of
entanglement revival can larger than that of before dropping.

Now we discuss the conditions of existing revival and larger revival. From
Eq.(8), we see if $\xi \geqslant \eta -J_{z},$ \ i.e., $b_{c}\geqslant \sqrt{%
(\eta -J_{z})^{2}-J^{2}}$, the concurrence $C$ will exhibit revival
phenomenon. But if we need larger revival value, combining $\frac{J}{\xi }>%
\frac{J\gamma }{\eta }$ and $\xi \geqslant \eta -J_{z}$, we have $J_{z}>\eta
-\frac{\eta }{\gamma }$. Therefore, the condition existing larger revival is
as 
\begin{equation}
\left\{ 
\begin{array}{c}
b_{c}\geqslant \sqrt{(\eta -J_{z})^{2}-J^{2}} \\ 
J_{z}>\eta -\frac{\eta }{\gamma }.%
\end{array}%
\right.
\end{equation}

\begin{figure}[tbp]
\includegraphics[width=2.8in, height=2.5in]{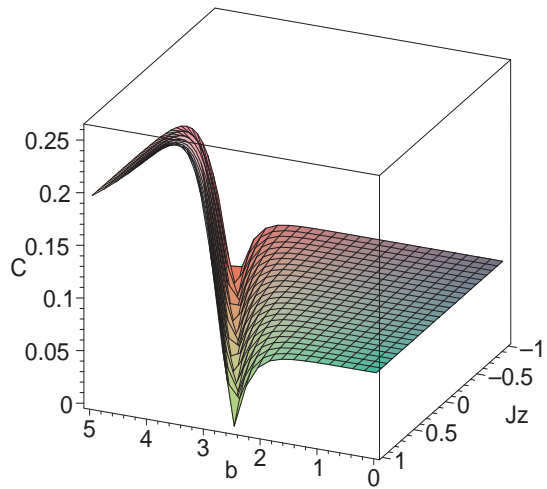}
\caption{(color online) Concurrence versus the inhomogeneous magnetic field
field $b$ and $J_{z}$, where $J=1$, $B=4$, $T=0.2$ and $\protect\gamma =0.3$%
. }
\end{figure}

In figure $2$ we give the plot of concurrence as a function of $b$ and $%
J_{z} $ at a fixed temperature and magnetic field for $\gamma =0.3$. It is
shown that the critical inhomogeneous magnetic field $b_{c}$ is increased
with the decrease of the interaction of $z$-component $J_{z}$ and there is
no entanglement at this critical point. As $b$ increases, the concurrence $C 
$ is initially constant which is nearly invariant by changing the
interaction of $z$-component $J_{z}$. It then decreases suddenly as the
critical inhomogeneous magnetic field $b_{c}$ is reached. However, instead
of vanishing for $b>b_{c} $, the concurrence persists and undergoes a
revival before decreasing. The most important thing is that the value of
entanglement revival can larger than that of before dropping. The maximal
entanglement value exists in the revival region, which means we still can
observe larger revival phenomenon for thermal entanglement.

\begin{figure}[tbp]
\includegraphics[width=2.80in, height=2.50in]{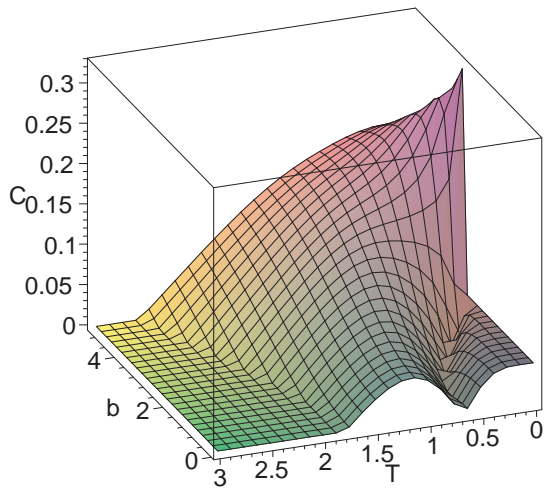}
\caption{(color online) Concurrence versus temperature and inhomogeneity for 
$\protect\gamma =0.2$ and $B=4$. The coupling constant $J=1$ and $J_{z}=1$.}
\end{figure}

With $B=4$, the concurrence as a function of $b$ and $T$ are given in Fig.3.
If $b$ is lower than a certain value, there are two areas showing
entanglement, and the entanglement first decreases and then undergoes a
revival before decreasing to zero. But when $b$ is larger than this certain
value, there is no revival phenomenon and the entanglement is decreased
monotonously with the increase of $T$. However, in the region of larger $%
(b,T) $ parameter space, we can derive larger entanglement. In addition, a
critical temperature $T_{c}$ is improved with the increase of $b$. The
property opens out the role of inhomogeneous magnetic field $b$ in improving
entanglement. Ref.[10] pointed out that a pairwise entanglement in $N$-qubit
isotropic Heisenberg system in certain degree can be increased by
introducing external field $B$. Our study shows the inhomogeneous magnetic
field $b$ can have a better action on improving entanglement for thermal
entanglement.

\begin{figure}[tbp]
\includegraphics[width=3.0in, height=3.0in]{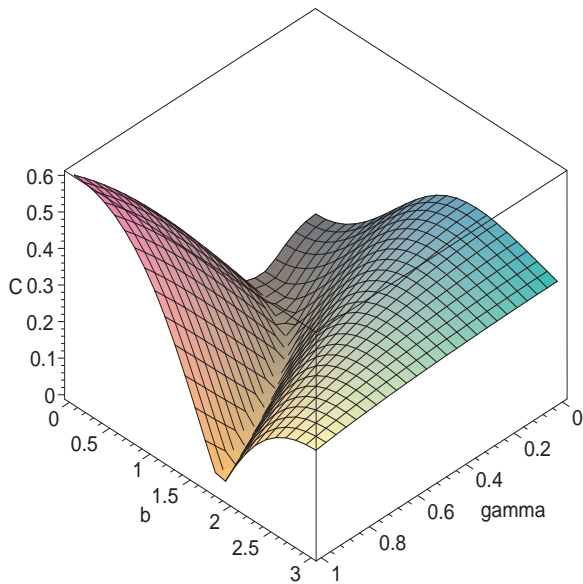}
\caption{(color online) The concurrence in the $XYZ$ spin model is plotted
vs $b$ and anisotropy parameter $\protect\gamma $, where $T=0.4$ and $B=0.8$%
. The coupling constant $J_{z}=-0.6$ and $J=1$.}
\end{figure}

Figure $4$ shows the entanglement as measured by the concurrence in terms of
the inhomogeneity $b$ and anisotropy parameter $\gamma $ at a finite
temperature $T$ for a finite external magnetic field $B$ $(B=0.8)$. We can
observe that entanglement exists in two regions: the one is in large $\gamma 
$ and small $b$; the other is in small $\gamma $ and $b$. If the anisotropy
of $XY$ plane $\gamma $ is large enough, we just need small inhomogeneous
external field $b$ or even without inhomogeneity, vice versa. At this point,
the role of $b$ is something similar to the anisotropy $XY$ plane. It seems
to be possible to control the anisotropy by adjusting external magnetic
field.

\section{Conclusion}

In conclusion, we have studied the thermal entanglement in an anisotropic
two qubits Heisenberg $XYZ$ chain under an inhomogeneous magnetic field.
Through analyzing the $T=0$ case, we find that conditions of existing
revival phenomenon and larger revival. The larger revival phenomenon still
exists at fixed finite temperature. If the parameters are proper and the
inhomogeneity is large enough, in the revival region the entanglement is
larger than that of without inhomogeneous magnetic field for fixed
temperature. Then we show by increasing the inhomogeneous magnetic field $b$
the entanglement and the critical temperature $T_{c}$ can be improved.

This work was supported by Natural Science Foundation of China under Grant
No. 10575017 and Natural Science Foundation of Liaoning Province of China
under No. 20031073.

\end{document}